 \documentclass[final,5p,times,twocolumn]{elsarticle}

 \usepackage{graphics}

\usepackage{amssymb}
\usepackage{amsmath}

\begin{document}

\begin{frontmatter}



\title{Quantum ratchet driven by broadband perturbation}


\author{D.V. Makarov}
\author{L.E. Kon'kov}

\address{
Laboratory of Nonlinear Dynamical Systems,\\
Pacific Oceanological Institute of the Russian Academy of Sciences,\\
43 Baltiiskaya st., 690041 Vladivostok, Russia, URL: http://dynalab.poi.dvo.ru}  

\begin{abstract}

Motion of an atomic ensemble trapped into a deep optical lattice is considered.
We propose a novel approach to construct an atomic ratchet
by superimposing two additional lattices whose amplitudes are small and subjected
to broadband modulation.
The broadband modulation is modeled by means of harmonic noise.
Directed atomic transport occurs with the properly chosen phase shift
between the signals modulating the amplitudes of the additional lattices.
It is shown that efficiency of the ratchet depends non-monotonously
on the parameter determining the spectral width of modulating signals.

\end{abstract}


\end{frontmatter}




\section{Introduction}
\label{Intro}

The ratchet effect, i.~e. occurrence of directed transport in space-periodic
systems under unbiased forcing,
has become an object of extensive research 
in recent years.
Quantum ratchets realized with cold atoms in optical lattices 
represent a promising way for constructing new-age microscopic devices,
for example, atomic chips in quantum computers.
In the simplest setup for the atomic ratchet, one uses 
far-detuned deep optical lattice \cite{Schiavoni,Jones04,Renzoni,Gommers05,Gommers08,Brown08}.
In this case atomic dynamics
can be fairly described in the framework of the semiclassical approximation.
Activation of net current requires escaping of atoms from the potential wells.
This process is impeded by the classical dynamical barriers in phase space.
Therefore, the experiments are usually conducted with relatively strong perturbation
of the optical potential, when the barriers don't survive.
Another approach to facilitate escaping
is the usage of high-frequency driving leading to the lowering 
of the effective optical potential \cite{Wickenbrock}.

In the present paper we propose a simple way to generate
directed atomic transport using a small-amplitude perturbation, 
even if atoms are initially localized
near the minima of the potential.
Usage of a small-amplitude perturbation should be favorable
because it allows one to avoid overheating of the atomic ensemble.
The key idea of our approach is to combine deterministic and stochastic driving.
The stochastic part is responsible for the destruction of the dynamical barriers, while
the deterministic one provides controllable violation of space-time symmetries,
which is necessary for production of atomic current in a desired direction.
In addition, a properly constructed deterministic driving allows one
to reduce the noise level needed for destruction of the dynamical barriers.
The particular scheme we consider is based on the usage 
of two additional optical lattices
whose amplitudes are subjected to broadband modulation.

The paper is organized as follows.
In the next section we describe the model considered.
Section \ref{Classical} is devoted to the properties of this model
in the classical limit.
Quantum atomic dynamics is considered in section \ref{Transport}.
In Summary, we give a brief account of the results obtained.

\section{The model}

In the case of large detuning of the atom-field resonance and in
the rotating-wave approximation, motion of a wavepacket corresponding
to the two-level atom is described by the Schr\"odinger equation
\begin{equation}
 i\hbar\frac{d\Psi}{dT}=-\frac{\hbar^2}{2m}\frac{\partial^2\Psi}{\partial X^2}
 +\hbar \frac{\Omega^2(X,T)}{\delta}\Psi,
\label{shrod0}
\end{equation}
where $X$ is position, $m$ is the atomic mass, 
$\delta$ is the detuning of the atom-field resonance, $\Omega$ is the Rabi frequency
which can be expressed as 
\begin{equation}
 \Omega^2(X,T)=\Omega_{\max}^2U(X,T).
\end{equation}
%
After the transformation
\begin{equation}
t=2\Omega_{\max}\sqrt{\frac{\omega_{\mathrm{r}}}{\delta}}T,\quad x
=2kX,\\
\label{norm2}
\end{equation}
where $k$ is the wavenumber of the laser field,
$\omega_{\mathrm{r}}=\hbar k^2/2m$ is the recoil frequency,
we rewrite 
the Schr\"odinger equation \cite{CCT11} in the following form:
\begin{equation}
 i\hbar_{\mathrm{resc}}\frac{\partial\Psi}{\partial t}=-\frac{\hbar_{\mathrm{resc}}^2}{2}\frac{\partial^2\Psi}{\partial x^2}
 + U(x,t)\Psi,
 \label{shrod}
\end{equation}
where the rescaled Planck constant is given by
\begin{equation}
 \hbar_{\mathrm{resc}}=4\sqrt{\delta\omega_{\mathrm{r}}}/\Omega_{\max}
\end{equation}
Hereafter we shall omit the subscript at $\hbar_{\mathrm{resc}}$.

In the present work, we propose the following configuration of the optical potential:
\begin{equation}
 U = 1-\cos{x} + \varepsilon V(x,t),
 \label{U}
\end{equation}
\begin{equation}
V(x,t) = 1 + f(t)\sin{x} - sf(t+\Delta)\cos{x},
\label{Vxt}
\end{equation}
where $\varepsilon\ll 1$, $s$ is a model parameter equal to 1 or $-1$,
and $f(t)$ is a broadband signal. We model this signal as the 
so-called harmonic noise \cite{HN,Anischenko}.
Harmonic noise is the two-dimensional Ornstein-Uhlenbeck process obeying
the following coupled stochastic differential equations
\begin{equation}
 \dot f=y,\quad
 \dot y=-\Gamma y-\omega_0^2f + \sqrt{2\mu\Gamma}\xi(t),
 \label{ou2d}
\end{equation}
where $\Gamma$ is a positive constant, and
$\xi(t)$ is Gaussian white noise with
\begin{equation}
\left<\xi(t)\right>=0,\quad
\left<\xi(t)\xi(t')\right>=\delta(t-t').
 \label{whitenoise}
\end{equation}
It should be emphasized that the terms $f(t)$ and $f(t+\Delta)$ in (\ref{Vxt})
correspond to one and the same realization of harmonic noise,
they differ only by the temporal shift $\Delta$.

The first two moments of harmonic noise are given by
\begin{equation}
 \left<f\right>=0,\quad
 \left<f^2\right>=\frac{\mu}{\omega_0^2}.
\end{equation}
Power spectrum of the harmonic noise is described by formula
\begin{equation}
 S(\omega)=\frac{\mu\Gamma}{\omega^2\Gamma^2 + (\omega^2-\omega_0^2)^2}.
 \label{spectrum}
\end{equation}
In the present work $\mu$ is taken of 1. Then the perturbation
strength is determined by the parameter $\varepsilon$.
In the case of low values of $\Gamma$,
the power spectrum has the peak at the frequency
\begin{equation}
 \omega_{\mathrm{p}}=\sqrt{\omega_0^2-\frac{\Gamma^2}{2}}
\label{w_p}
\end{equation}
with the width
\begin{equation}
\Delta\omega = \sqrt{\omega_{\mathrm{p}}+\Gamma\omega'}-
\sqrt{\omega_{\mathrm{p}}-\Gamma\omega'},
 \label{width}
\end{equation}
where $\omega'=\sqrt{\omega_0^2-\Gamma^2/4}$.
It should be mentioned that harmonic noise had been recently
considered in the context of the Landau-Zener tunneling
in optical lattices \cite{Wimberger-NJP,Wimberger-FNL}.

\begin{figure}[!htb]
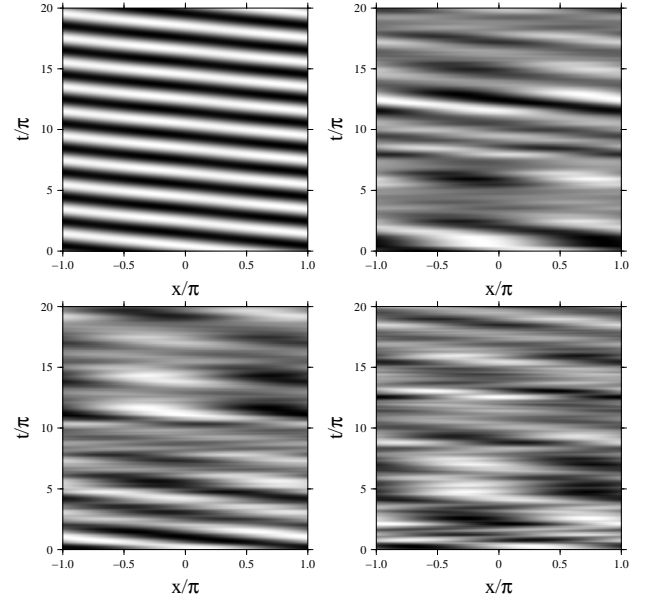

\begin{center}
\includegraphics[width=0.22\textwidth]{fig1-1.eps}
\includegraphics[width=0.22\textwidth]{fig1-2.eps}\\
\includegraphics[width=0.22\textwidth]{fig1-3.eps}
\includegraphics[width=0.22\textwidth]{fig1-4.eps}
\caption{
Patterns of the perturbation $V(x,t)$ in the $x-t$ plane.
Left upper panel corresponds to the  deterministic case $\Gamma=0$.
Other panels correspond to typical realizations of $V(x,t)$
for nonzero values of $\Gamma$:
$\Gamma=0.5$ (right upper panel),$\Gamma=1.0$ (left lower panel), and $\Gamma=1.5$
(right lower panel). All plots correspond to $s=1$.
}
\label{fig-V}
\end{center}
\end{figure}

As $\Gamma\to 0$, 
$f(t)\to \sin(\omega_0t+\phi_0)$, where $\phi_0$ is determined
by initial conditions in (\ref{ou2d}).
Setting $f(0)=1$, $y(0)=0$, and 
\begin{displaymath}
 \Delta=\frac{\pi}{2\omega_0},
 \end{displaymath}
one obtains
$f = \cos{\omega_0t}$ and
\begin{equation}
V(x,t)\to \sin(x+s\omega_0t),
\label{plane}
\end{equation}
with $\Gamma\to 0$. Hence, it turns out that $V(x,t)$
for $\Gamma>0$ behaves like a plane wave with 
fluctuating phase and amplitude (see Fig.~\ref{fig-V}).
In numerical simulation, we set $\omega_0=1$ and $\varepsilon=0.05$.

Taking into account the periodicity of the potential $U$,
we use the periodic boundary conditions
\begin{equation}
\Psi(x) = \Psi(x+2\pi).
 \label{BC}
\end{equation}
It means that we consider only the states with zero quasimomentum.
This choice is reasonable as long as we consider
only the semiclassical regime with
fairly small values of the rescaled Planck
constant $\hbar$.
In this case energy bands are flat
and tunneling between neighboring potential 
wells is negligible, that is, a wave function of a single
atom is tightly enough localized in space.

\section{Classical dynamics}
\label{Classical}

\begin{figure}[!htb]
\begin{center}
\includegraphics[width=0.48\textwidth]{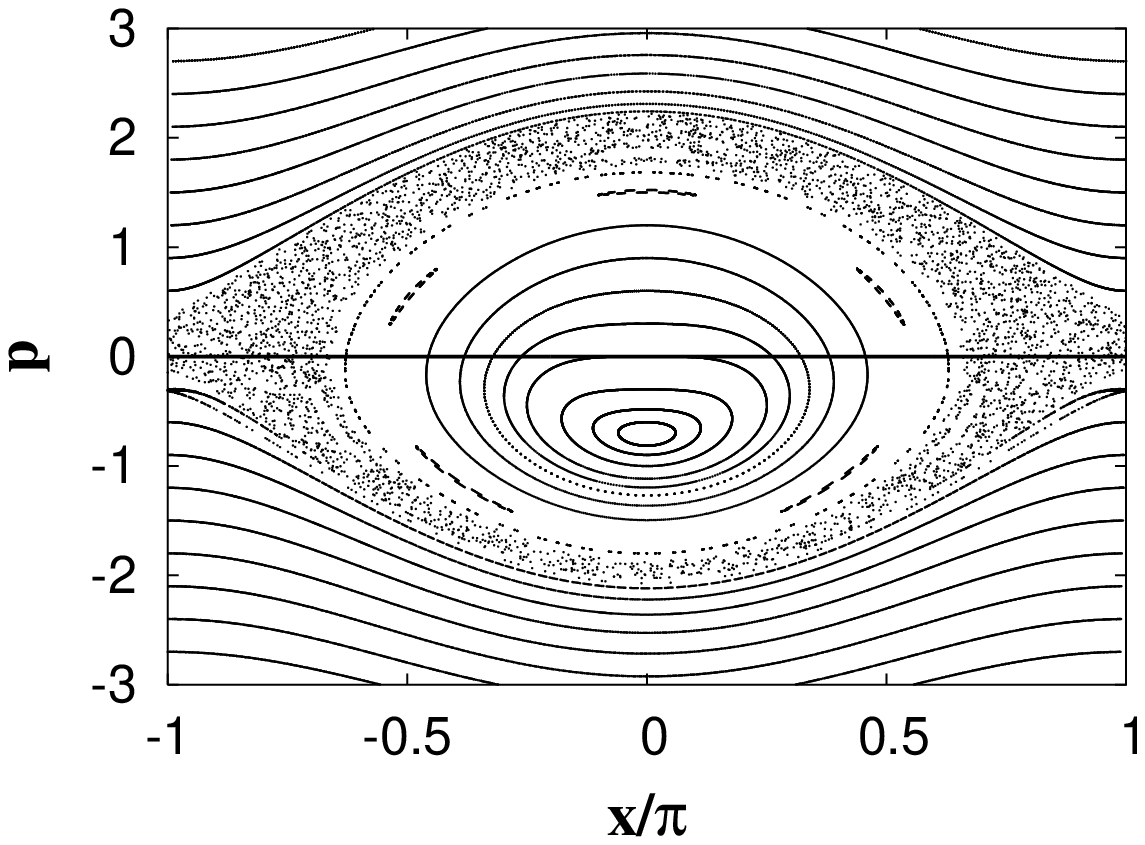}\\
\includegraphics[width=0.48\textwidth]{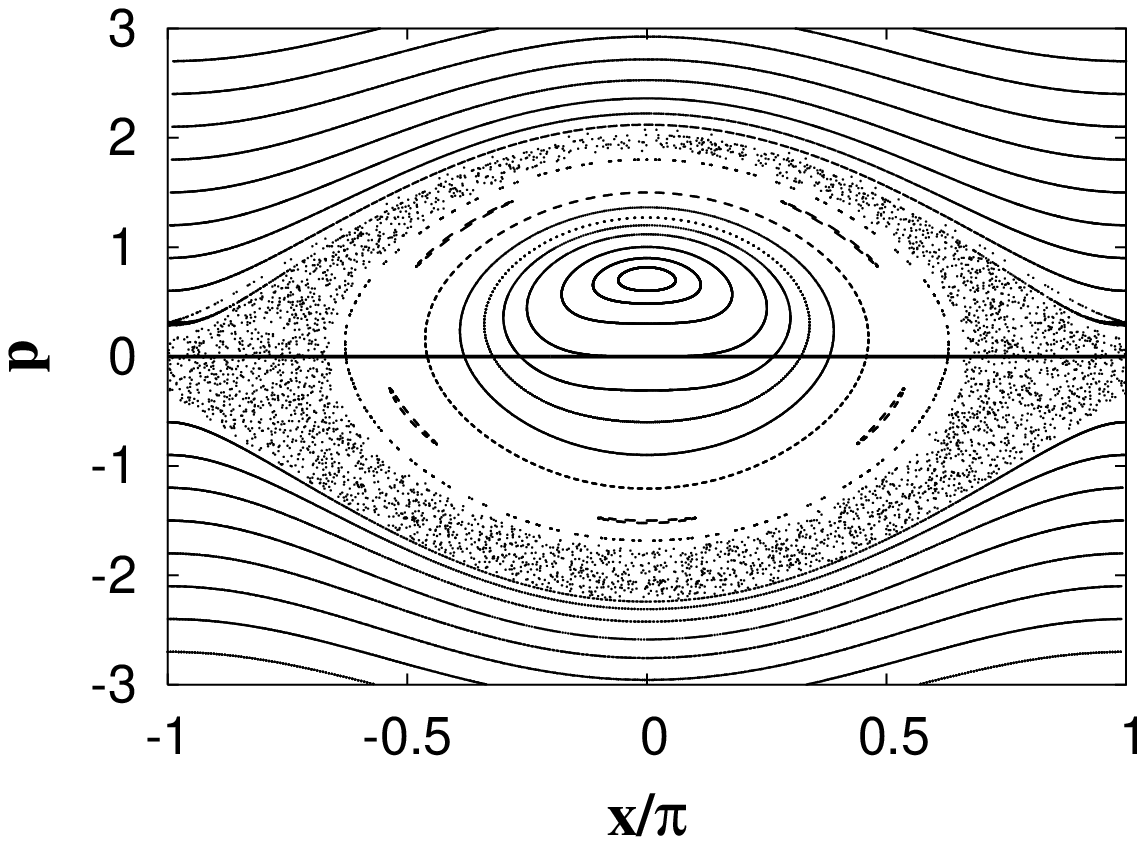}
\caption{
Poincar\'e sections for the deterministic case $\Gamma=0$.
Upper and lower plots correspond to $s=-1$ and $s=1$, respectively.
Horizontal lines in the plots correspond to $p=0$.
}
\label{Fig-poinc}
\end{center}
\end{figure}

Before considering quantum evolution of the atomic ensemble,
it is reasonable to give insight onto atomic dynamics in the 
classical limit.
Trajectories of atoms obey  Hamiltonian equations of motion
\begin{equation}
\frac{dx}{dt}=\frac{\partial H}{\partial p},\quad
\frac{dp}{dt}=-\frac{\partial H}{\partial x},
 \label{sys}
\end{equation}
where the classical Hamiltonian $H$ is given by 
\begin{equation}
 H = \frac{p^2}{2} - \cos{x} + \varepsilon V(x,t).
\end{equation}
Time-dependent perturbation $V(x,t)$ leads to the nonintegrability 
of equations (\ref{sys}) and onset of chaos.
For $\Gamma=0$ and $\varepsilon\ll 1$, chaotic motion 
occurs only inside a relatively small near-separatrix layer acting as a ``bridge''
between the domains of finite and ballistic motion.
In this case, 
the perturbation is a plane wave of the form (\ref{plane}),
and the equations of motion read
\begin{equation}
\frac{dx}{dt}=p,\quad
\frac{dp}{dt}=-\sin{x}-\varepsilon\cos(x+s\omega_0 t).
 \label{sys-det}
\end{equation}
 One can easily ensure that both the shift symmetry
\begin{equation}
 x \to -x,\quad t \to t+\frac{\pi}{\omega_0},
\end{equation}
and the time-reversal symmetry
\begin{equation}
  x \to -x,\quad t \to -t,
\end{equation}
are broken in this case, making the onset of directed transport possible.
The plane wave $V(x,t)$
tends to carry atoms in the direction $x\to-\infty$ ($x\to\infty$) for $s=1$ ($s=-1$),
i.~e. along the lines of the of the constant phase in Fig.~\ref{fig-V}.
It results in the asymmetry of the chaotic layer in the momentum space,
as it is demonstrated in Fig.~\ref{Fig-poinc}.
This asymmetry expects emergence
of the nonzero ballistic flux \cite{JETPL,PRE75,JTPL08,EPJB} whose
direction is determined by $s$.
However, there is a large layer of regular motion in classical
phase space, acting as a dynamical barrier for the atoms localized near the potential minima and
preventing their transition into the ballistic regime.
Consequently, atoms with minimal initial energies
can produce directed current only due to tunneling which 
is almost negligible in deep optical lattices.

With $\Gamma>0$, the perturbation
contains a noisy component which 
causes atom diffusion through the dynamic barriers.
The correlation time of the noisy component
can be roughly estimated as $\Delta\omega^{-1}$, therefore,
increasing of $\Gamma$ should anticipate enhancing of the diffusion.
The noisy component doesn't alter time-space symmetries, therefore,
the direction of the resulting transport is still determined by $s$.
Indeed, one can see in Fig.~\ref{fig-V} that the constant phase lines
for $V(x,t)$ are biased in the same direction for all values of $\Gamma$, 
despite of the gradual pattern randomization with increasing $\Gamma$.

It should be noted that the stable equilibrium point in the Poincar\'e
plots presented in Fig.~\ref{Fig-poinc} is displaced from the origin $x=0, p=0$.
This occurs due to nonlinear resonance of multiplicity 1:1 between oscillations of 
the perturbation with the frequency $\omega_0=1$, and the unperturbed oscillations
in the vicinity of the original equilibrium point.
As the resonance 1:1 has relatively large width in phase space, the atoms 
it traps can undergo large-amplitude oscillations in the energy space, remaining
inside the regular domain though.
If $\Gamma>0$, the presence of such oscillations facilitates activation
of atoms initially located near the center of phase space, because the resonance 
acts as a ``lift'' carrying atoms from the center upwards in energy, thus reducing
their activation energy.
In this way, $\omega_0=1$ seems to be close to the optimal choice of the driving frequency \cite{Koshel-DAN, Koshel-Chaos}.
Indeed, it is well-known that the lowest-order resonances typically have the largest widths in the energy space,
thereby providing the most efficient excitation.


\section{Quantum transport}
\label{Transport}

Let's proceed with considering quantum evolution by solving
numerically the Schr\"odinger equation (\ref{shrod}).
Given a realization of the modulation $f(t)$,
one can calculate the asymptotic current defined as \cite{Flach07}
\begin{equation}
J(t) = \frac{1}{t}\int\limits_{0}^{t}p_{\mathrm{q}}(t')\,dt',
 \label{Jdef}
\end{equation}
where $p_{\mathrm{q}}$ is quantum-mechanical
momentum expectation value
\begin{equation}
 p_{\mathrm{q}}(t) = \int\,dx\, \Psi^{*}\hat p \Psi,\quad
 \hat p = -i\hbar\frac{\partial}{\partial x}.
\end{equation}
Averaging over realizations of $f(t)$, we obtain
the statistical mean of the asymptotic current
\begin{equation}
 \left<J\right> = \frac{1}{N}\sum\limits_{n=1}^N J^{(n)}.
\end{equation}
where integer $n$ labels the realizations.
Our computations were performed with $N=100$.
As long as we consider the case of a deep optical lattice,
the rescaled Planck constant $\hbar$ was taken of 0.1.
Initial state used is a random superposition of coherent states:
\begin{equation}
\begin{aligned}
 \Psi(t=0)&=A\sum\limits_{n=1}^{N_{\text{cs}}} \phi(x_0,p_n),\\
  \phi(x_0,p_n)&=(2\pi)^{-1/4}\Delta_x^{-1/2}e^{-\frac{(x-x_0)^2}{4\Delta_x^2} + \frac{ip_n(x-x_0)}{\hbar}},
\end{aligned}
\end{equation}
where $x_0=0$, $\Delta_x=0.25$, $N_{\text{cs}}=10^4$,
and $p_n$ is random quantities obeying the Gaussian statistics with zero mean and variance
0.25. This initial condition corresponds to a wavepacket tightly localized near the stable equilibrium point 
of the unperturbed system $x=0$, $p=0$.

\begin{figure}[!htb]
\begin{center}
\centerline{
\includegraphics[width=0.4\textwidth]{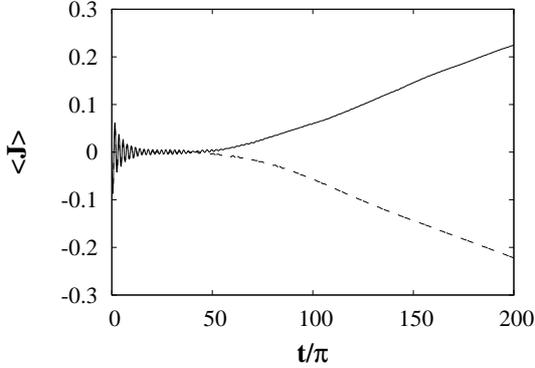}}
\caption{
Time dependence of the ensemble-averaged asymptotic current for $s=-1$ (solid)
and $s=1$ (dashes). In both cases $\Gamma=0.4$.
}
\label{fig-fluxes}
\end{center}
\end{figure}

Figure \ref{fig-fluxes} represents time dependence of the ensemble-averaged
current for $\Gamma=0.4$. 
One can see that the ballistic current is activated only beyond some time threshold
which is needed to atoms to gain enough energy for escaping.
Below this threshold, $\left<J\right>(t)$ undergoes the oscillations associated with atomic motion inside the potential wells.   
Notably, the curves corresponding
to $s=-1$ and $s=1$ are nearly symmetric with respect to the semiaxis $\left<J\right>=0$.
It means that the current direction is readily controlled by the sign of $s$.

\begin{figure}[!htb]
\begin{center}
\centerline{
\includegraphics[width=0.4\textwidth]{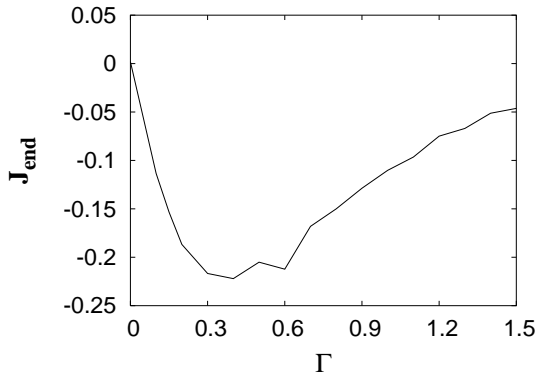}}
\caption{
Dependence of the ensemble-averaged asymptotic current at $t=200\pi$ 
on $\Gamma$.
}
\label{fig-flux_G}
\end{center}
\end{figure}

It is reasonable to examine how current varies with increasing of the noise contribution 
which can be quantified by $\Gamma$.
Indeed, small noise contribution anticipates weak diffusion and slow crossing of the dynamic
barriers in phase space.
On the other hand, large amount of noise should 
destroy the plane-wave form of the perturbation due to the loss of coherence between
the modulating signals $f(t)$ and $f(t+\Delta)$. 
As long as the plane-wave form is responsible
for violation of space-time symmetries, resulting transport should become undirected.
Consequently, there should be some 
intermediate range of $\Gamma$ values corresponding to the most efficient
activation of directed current.
To find it, we consider finite time interval $[0:200\pi]$
and calculate the dependence of the mean asymptotic current at the endpoint
of the interval on $\Gamma$.
The result computed with $s=1$ is presented in Fig.~\ref{fig-flux_G}. 
According to above expectations, the dependence of $J_{\mathrm{end}}$ on $\Gamma$
is not monotonic, and
the most efficient generation of ballistic current corresponds
to $0.3<\Gamma<0.6$.
It should be mentioned that directed transport is almost absent in the purely deterministic
case $\Gamma=0$. It unambiguously indicates on the importance of noise in the activation
of ballistic atomic current.


%
\begin{figure}[tpb]
\begin{center}
\centerline{
\includegraphics[width=0.4\textwidth]{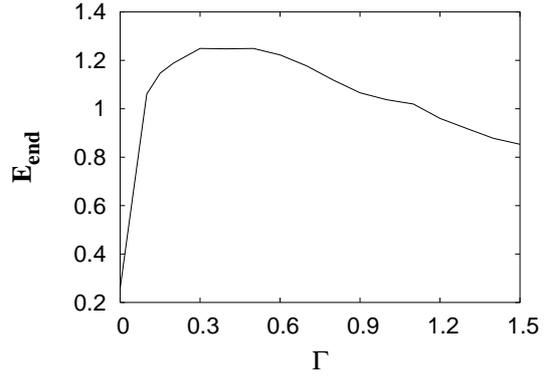}}
\caption{
Ensemble-averaged energy at $t=200\pi$ vs  $\Gamma$.
}
\label{fig-en_G}
\end{center}
\end{figure}

To underline the link between the onset of directed current
and heating of atoms, we calculate
the mean energy determined as
\begin{equation}
 \left<E\right>(t) = \frac{1}{N}\sum\limits_{n=1}^N \int\,dx\,\Psi^*(t)\hat H^{(n)}\Psi(t),
\end{equation}
where 
\begin{displaymath}
 \hat H^{(n)}=
 -\frac{\hbar^2}{2}\frac{\partial^2\Psi}{\partial x^2}+
 1 - \cos{x} + \varepsilon V^{(n)}(x,t)
\end{displaymath}
is the Hamiltonian operator corresponding to 
the $n$-th realization of perturbation $V(x,t)$.
Dependence of $E_{\mathrm{end}}\equiv\left<E\right>(t=200\pi)$ on $\Gamma$, demonstrated in
Fig.~\ref{fig-en_G}, reveals that the range of the $\Gamma$ values
corresponding to the most efficient current generation, $0.3\le\Gamma\le0.6$,
simultaneously corresponds to the most efficient heating.
Non-monotonous dependence on the noise level indicates on the significance
of factors of heating, which are not concerned with noise. 
In our case, the role of such factor is played by classical resonance 1:1 whose impact
decreases as $\Gamma$ grows.
Thus, the optimal activation occurs when resonance-assisted and diffusive mechanisms of atom heating 
accompany each other constructively.
It should be noted that decreasing of $E_{\mathrm{end}}$ for
$\Gamma>0.6$ is significantly slower than the corresponding decreasing of $|J_{\mathrm{end}}|$ 
(see Fig.~\ref{fig-flux_G}).
This can be understood as noise-induced recovery of space-time symmetries due to loss
of correlations between harmonic noise values
$f(t)$ and $f(t+\Delta)$.

\section{Summary}

We presented a novel approach to produce directed atomic current
in an optical lattice by means of weak unbiased perturbation
consisted of two lattices with broadband amplitude modulation.
In this work we model the broadband modulation as harmonic noise.
The approach presented allows for the current generation even if atoms are tightly
confined by the lattice potential at the initial moment.
It is shown that the efficiency of generation depends non-monotonously 
on the parameter describing the spectral width of the perturbation.
We suppose that the approach reported can be used for production of directed
current in a more complicated configurations of the optical potential, for example,
in potentials with random disorder.

\section*{Aknowledgments}

We acknowledge useful discussions with S.~Prants, D.~Maksimov,
A.~Kolovsky and M.~Uleysky.
This work was supported by the grants of the Russian Foundation of Basic Research
(projects 09-02-01258 and 12-02-31416), and the integration grant of the Far-Eastern and 
Siberian Branches of the Russian Academy of Sciences (project 12-II-07-022).







\end{document}